\documentclass[longauth]{aa} 

\usepackage{longtable}
\usepackage{graphicx}
\usepackage{txfonts}
\begin{document}

   \title{The Gaia-ESO Survey: the N/O abundance ratio in the Milky Way
   \thanks{Based on observations collected with the FLAMES instrument at
VLT/UT2 telescope (Paranal Observatory, ESO, Chile), for the Gaia-
ESO Large Public Spectroscopic Survey (188.B-3002, 193.B-0936).}}

   \author{L. Magrini\inst{1},   F. Vincenzo\inst{2}, S. Randich\inst{1}, E. Pancino\inst{1,3}, G. Casali\inst{1,4},  G. Tautvai\v{s}ien\.{e}\inst{5}, A. Drazdauskas\inst{5}, \v{S}. Mikolaitis\inst{5}, R. Minkevi\v{c}i\={u}t\.{e}\inst{5}, E. Stonkut\.{e}\inst{5}, Y. Chorniy\inst{5}, V. Bagdonas\inst{5},  G. Kordopatis\inst{6}, E. Friel\inst{7}, V. Roccatagliata\inst{1}, F. M. Jim\'{e}nez-Esteban\inst{8}, G. Gilmore\inst{9},   A. Vallenari\inst{10}, T. Bensby\inst{11},   A. Bragaglia\inst{12},  A.~J. Korn\inst{13},  A.~C. Lanzafame\inst{14},   R. Smiljanic\inst{15},    A. Bayo\inst{16,17}, A.~R. Casey\inst{18,19},   M.~T. Costado\inst{20},  E. Franciosini\inst{1},  A. Hourihane\inst{9}, Jofr\'e \inst{21},  J. Lewis\inst{9},   L. Monaco\inst{22},    L. Morbidelli\inst{1},  G. Sacco\inst{1},  C. Worley\inst{9} }

 \institute{INAF - Osservatorio Astrofisico di Arcetri, Largo E. Fermi, 5, I-50125 Firenze, Italy
\email{laura@arcetri.astro.it} \and 
Centre for Astrophysics Research, University of Hertfordshire, College Lane, Hatfield, AL10 9AB, UK\and
Space Science Data Center - Agenzia Spaziale Italiana, Via del Politecnico SNC, 00133 Roma\and
Dipartimento di Fisica e Astronomia, Universit\'a degli studi di Firenze, Via Sansone 1, I-50019 Sesto Fiorentino, Italy\and
Institute of Theoretical Physics and Astronomy, Vilnius University, Saul\.{e}tekio av. 3, 10257 Vilnius, Lithuania\and
Universit\'e C\^{o}te d'Azur, Observatoire de la C\^{o}te d'Azur, CNRS, Laboratoire Lagrange, France  \and
Department of Astronomy, Indiana University, Bloomington, IN, USA\and
Departamento de Astrof\'{\i}sica, Centro de Astrobiolog\'{\i}a (INTA-CSIC), ESAC Campus, Camino Bajo del Castillo s/n, E-28692 Villanueva de la Ca\~nada, Madrid, Spain\and
Institute of Astronomy, Madingley Road, University of Cambridge, CB3 0HA, UK\and
INAF - Osservatorio Astronomico di Padova, Vicolo dell'Osservatorio 5, 35122 Padova, Italy \and 
Lund Observatory, Department of Astronomy and Theoretical Physics, Box 43, SE-221 00 Lund, Sweden \and
INAF - Osservatorio Astronomico di Bologna, via Gobetti 93/3, 40129, Bologna, Italy \and
Department of Physics and Astronomy, Uppsala University, Box 516, SE-751 20 Uppsala, Sweden\and
Dipartimento di Fisica e Astronomia, Sezione Astrofisica, Universit\'a di Catania, via S. Sofia 78, 95123, Catania, Italy \and 
Nicolaus Copernicus Astronomical Center, Polish Academy of Sciences, ul. Bartycka 18, 00-716, Warsaw, Poland\and
Instituto de F\'isica y Astronomi\'ia, Universidad de Valparai\'iso, Chile\and
N\'ucleo Milenio Formaci\'on Planetaria - NPF, Universidad de Valpara\'iso, Av. Gran Breta\~na 1111, Valpara\'iso, Chile\and
  Monash Centre for Astrophysics, School of Physics \& Astronomy, Monash University, Clayton 3800, Victoria, Australia\and
  Monash Faculty of Information Technology, Monash University, Clayton 3800, Victoria, Australia\and
  Departamento de Did\'actica, Universidad de C\'adiz, 11519 Puerto Real, C\'adiz, Spain\and
  N\'ucleo de Astronom\'{i}a, Facultad de Ingenier\'{i}a, Universidad Diego Portales, Av. Ej\'ercito 441, Santiago, Chile\and
  Departamento de Ciencias Fisicas, Universidad Andres Bello, Fernandez Concha 700, Las Condes, Santiago, Chile
 }
    \date{Received ; accepted }

 
  \abstract
   {The abundance ratio N/O is a useful tool to study the interplay of galactic processes, e.g. star formation efficiency, time-scale of infall and outflow loading factor }
   {We aim to trace $\log$(N/O) versus [Fe/H] in the Milky Way and to compare it with a set of chemical evolution models to understand the role of infall, outflow and star formation efficiency in the building-up of the Galactic disc.    }
   {We use the abundances from {\sc idr2-3, idr4, idr5} data releases of the Gaia-ESO Survey both for  Galactic field and open cluster stars. 
   We determine membership and  average composition of open clusters and we  separate thin and thick disc field stars.  
   We consider the effect of mixing in the abundance of N in giant stars. We compute a grid of chemical evolution models, suited to reproduce the main features of our Galaxy, exploring the effects of the star formation efficiency, the infall time-scale and the differential outflow.  }
   {With our samples, we map the metallicity range -0.6$\leq$[Fe/H]$\leq$0.3 with a corresponding -1.2$\leq$$\log$N/O$\leq$-0.2, where the secondary production of N dominates.
   Thanks to the wide range of Galactocentric distances covered by our samples, we can distinguish the behaviour of $\log$(N/O) in different parts of the Galaxy.     }
   {
   Our spatially resolved results allow us to distinguish differences in the  evolution of N/O with Galactocentric radius. 
   Comparing the data with our models, we can characterise the radial regions of our Galaxy. A shorter infall time-scale is needed in the inner regions, while the outer regions need a longer infall time-scale, coupled with a  higher star formation efficiency.
   We  compare our results with nebular abundances obtained in MaNGA galaxies, finding in our Galaxy a much wider range of $\log$(N/O)  than in integrated observations of external galaxies of similar stellar mass, but similar to the ranges found in studies of individual H~{\sc ii} regions. 
    }

\authorrunning{Magrini, L. et al.}
\titlerunning{N/O in Open Clusters and Galactic discs }

   \maketitle
%

\section{Introduction}
Nitrogen, one of the most common elements in the Universe and one of the key-ingredients at the basis of life as we know it \citep[e.g.][]{sa16},  has a complex nucleosynthesis \citep[see, e.g,][]{vincenzo16, vincenzo18}. 
It is mostly produced by low- and intermediate-mass stars (LIMS) with metallicity dependent yields. 
The metallicity dependence in the production of N is related to its double nuclear channels:  the so-called primary and secondary productions. 
The primary component directly derives from the burning of H and He and  it does not require any previous enrichment in metals.  
In the LIMS, this component is produced during the third dredge-up in  the asymptotic giant branch (AGB) phase \citep[see, e.g.][]{RV81, vandenHoekGroenewegen97, liang01,henry00}.
On the other hand, the secondary N component increases with metallicity since it is related to the CNO cycle, 
in which N is formed using previously produced C and O. 
However, the primary and secondary productions of N in LIMS are not enough to reproduce 
the observed plateau in N/O observed at very low metallicities.  Including the production of N in massive low-metallicity stars might help, although our knowledge of N stellar yields for massive stars is still inadequate \citep[cf.][]{MM00,MM02A, MM02B, CL04, CL13,gil13, taka14}.  N abundances in stars, together with C abundances, are extremely  useful for Galactic astro-archaeology studies because the observed C/N ratio in evolved stars has been shown to correlate well with the age of the stars 
\citep[see, e.g.][]{salaris2015,marting2016,masseron17,feuillet2018, casali18}. 

From an observational point of view,  the abundance ratio N/O in galaxies  is usually measured through emission-line spectroscopy of HII regions or of unresolved star-forming regions, both by individual studies \citep[see, e.g.][]{vila92, vanzee98, vanzee06, perez09, izotov12, berg12, james15,kumari2018} and by large surveys, 
as for instance the Sloan Digital Sky Survey (SDSS) \citep{Sloan} and the Sloan Digital Sky Survey IV Mapping Nearby Galaxies at Apache Point Observatory survey (MaNGA), from which N/O was estimated in a large number of star-forming galaxies  \citep[cf.][]{liang06, bundy15, vincenzo16,belfiore17}. 
The basic trend found collecting extragalactic datasets is {\em i)} 
a significant positive slope of N/O  versus oxygen abundance in the metal-rich regime, related to the secondary production and {\em ii)} a plateau of N/O for  low-metallicity galaxies. 
The two metallicity regimes can be divided approximately at 12 + $\log$(O/H)$\sim$8~dex \citep[see, e.g.][]{henry00, contini02}.  Finally, there are several studies based on Galactic samples of stars and of H~{\sc ii} regions designed to study the evolution of nitrogen in our Galaxy \citep[e.g.][]{israelian04, christlieb04, spite05,esteban05, carigi05,rudolph2006,esteban2018};  these measurements of N and O abundances in stars are particularly important to compare with our samples (see Sect.~5).

N and O abundances can be measured in absorption with very high resolution in the interstellar medium (ISM) of galaxies lying along the line of sight to quasars (namely, in the so-called Damped Ly$\alpha$ systems); examples of studies in this sense are \citet{pettini2002,pettini2008} and \citet{zafar2014} (but see also \citealt{vangioni2018} for a more theoretical point of view). At high redshifts, the N/O ratio can also be measured from the analysis of the spectra of galaxies hosting an active galactic nucleus, a supernova or a gamma ray burst, by making use of detailed numerical codes taking into account the photoionisation and shock of the ISM \citep[see, e.g.][]{contini2015,contini2016,contini2017a,contini2017b,contini2018}.

The interpretation of the origin and evolution of nitrogen in our Galaxy was faced by \citet{chiappini05}, making use of N/O measured in low-metallicity stellar spectra \citep{israelian04, christlieb04, spite05}; in particular, they compared the observed abundances in our Galaxy with the so-called "two-infall model'', as originally developed by \citet{chiappini97, chiappini01}, according to which the Galaxy was assembled from two separated episodes of gas accretion with different typical timescales (the shortest giving rise to the halo and thick disc of our Galaxy, the longest to the thin disc), 
in which they varied the assumed sets of stellar nucleosynthetic yields to understand the origin of the low-metallicity plateau in N/O. 
A similar analysis was done by \citet{gavillan06}, who made use of a collection of Galactic data sets of stellar and nebular abundances,
to constrain their chemical evolution models.  To understand  the historical problem of nitrogen evolution, they introduced a 
primary component in intermediate-mass stars. In addition, they explained the dispersion of N/O to be due to a variation of the star formation rates
across the Galactic disc. This was also confirmed by \citet{molla06} who analysed the role played by star formation efficiency  in the evolution of the abundance ratio N/O.

With the advent of large spectroscopic surveys, such as Gaia-ESO \citep{gilmore12}, APOGEE \citep{apo} and GALAH \citep{galah}, that allow the measurement 
of a large variety of elements in stars of the Milky Way, it is possible to investigate with sizeable statistical samples the behaviour of N/O in the different populations of our Galaxy.
Among the several on-going spectroscopic surveys, the Gaia-ESO Survey \citep[GES, ][]{gilmore12,RG13} has provided high resolution spectra of 
different stellar populations of our Galaxy using the spectrograph FLAMES@VLT \citep{pasquini02}. 
GES aims at homogeneously deriving stellar parameters and abundances in a large variety of environments, including the major Galactic components (thin and thick discs, halo, bulge), 
open and globular clusters and calibration samples. 
The higher resolution spectra obtained with UVES allow the determination of abundances of more than 30 different elements, including also nitrogen and oxygen both in field and in cluster stars. 

In the near future, astronomers will couple the detailed chemical abundance information from all the aforementioned Galaxy spectroscopic surveys with precise spatial and kinematical information of the stars as provided by \textit{Gaia} \citep{lindegren2016,brown2018} and stellar age information from asteroseismology studies \citep[see, e.g.][]{casagrande2014,casagrande2016}, to study the velocity and density fields as drawn by stars with different ages, location  and chemical abundances in the Galaxy. 

The paper is structured as follows: in Section~2 we describe the spectral analysis and in Section~3 we present our samples. In Section~4 we discuss the effect of mixing on nitrogen abundances in evolved stars.
In Section~5 we describe the set of chemical evolution models adopted to compare with the data, and in Section 6  we give our results and in Section 6 we compare the Milky Way with Local Universe results.  In Section~8 we give our summary and conclusions. 

\section{Spectral analysis}
The abundance analyses of N and O were performed by one of the Nodes of GES (Node of Vilnius), in the Working Group (WG) analysing the UVES spectra (WG~11); their derived abundances are among the recommended products of GES.  

In the optical spectral range, the abundances of N and O are derived from molecular bands and atomic lines of these elements, in some cases combined with carbon.  
In particular, in the analysis of the optical stellar spectra, the $^{12}{\rm C}^{14}{\rm N}$ molecular bands in the spectral range $6470-6490$~\AA, the ${\rm C}_2$ Swan (1,0) band head at $5135$~\AA, the ${\rm C}_2$ Swan (0,1) band head at $5635.5$~\AA, and the forbidden [O~{\sc i}] line at $6300.31$~\AA\,  are used; all these molecular bands and atomic lines are analysed through spectral synthesis with the code BSYN \citep{tau15}. 

To derive N and O abundances, all these lines and bands are analysed simultaneously; in this iterative process, also the determination of the C abundance is included.
For the determination of the oxygen abundance, we take into account the oscillator strengths of the two lines of\textsuperscript{58}Ni and \textsuperscript{60}Ni that are blended with the oxygen line \citep{johansson03}. 
The synthetic spectra are calibrated on the solar spectrum from \citet{kurucz05}, with the solar abundance scale of \citet{grevesse07} to make the analysis strictly differential. 
We adopt the MARCS grid of model atmospheres \citep{gustafsson08}.
In the fitting procedure of the observations with theoretical spectra, we take into account stellar rotation, which is one of the products 
of GES; the measurements of stellar rotation are described in \citet{sacco14}.

The atmospheric parameters of the stars are spectroscopically determined combining the results of several Nodes, with a methodology described in \citet{smi14}. Average uncertainties in the atmospheric parameters are 55~K, 0.13~dex, and 0.07~dex
for $T_{\rm eff}$, log~$g$, and [Fe/H], respectively.
The uncertainties on the abundances of N and O are estimated considering the errors on the atmospheric parameters and random errors, which are mainly caused by uncertainties of the continuum placement and by the signal-to-noise (S/N). 
We also take into account in the error budget the interplay between abundance of C, N and O in the simultaneous determination of their abundances.  
Considering all these aspects, typical errors on nitrogen and oxygen abundances are $\sim$0.10 and $\sim$0.09~dex, respectively. More details about the method of analysis and the evaluation of the uncertainties are given in \citet{tau15}.

\section{The samples}
Our samples are composed by Milky Way field stars and stars in open clusters. 

The former sample includes  stars observed with the UVES setup centred around 580.0 nm that belong to the  solar neighbourhood and  the inner disc samples.  
We have divided the field stars into thin and thick disc stars, using their [$\alpha$/Fe] abundance ratio, following the approach of   \citet{adibekyan11}.  
Our sample includes 19 thin disc stars and 130 thick disc stars with both N and O measurements. 
Their stellar parameters and the abundances of N and O  are listed in Table~\ref{tab_stars} in Appendix A. 

The number of stars in our samples depends on the observed number of dwarfs and giants and on their S/N as well.  This is because the CN molecular bands are less pronounced in dwarfs and not always they can be measured.  
In dwarf stars, the typical depth of the largest CN molecular band is $\sim$0.02 (in relative intensity with respect to the continuum) and, even at high S/N$\sim$100,  there are fluctuations of about 0.01 in this feature. The other features are even smaller. Consequently, to analyse them in dwarf stars, 
we need spectra with S/N as high as $\sim$200, which was not always achieved.  
On the other hand, CN bands in giant stars are much larger for two main reasons. 
The former is that molecular lines are larger in stars with lower temperatures, thus we can determine accurate abundances from spectra with lower S/N. 
The latter is because giant stars can be enriched in N due to internal chemical mixing processes.  
Therefore, there are far more determinations of nitrogen for giant stars compared to the stars in earlier stages of their evolution. 

For the above reasons, and because the GES selection function \citep{sel} favours giant stars in the thick disc and dwarf stars in the thin disc, 
in our samples we are biased towards thick-disc stars.   

The latter sample includes open clusters  with ages$>$0.1~Gyr whose parameters and abundances have been delivered  in the four Gaia-ESO releases {\sc idr2, idr3, idr4, idr5}. 
Membership analysis has been performed as in \citet{magrini18},  using the [Fe/H] abundance and radial velocity distributions to define cluster members. 
Most of the member stars in open clusters are giant stars, belonging to the Red Clump (RC) phase. 
In Table~\ref{tab_clu},  we present the main cluster parameters (age, turn-off stellar mass, distance, metallicity [Fe/H]), the median abundances of oxygen and nitrogen with their standard deviations,  the number of stars and the reference data release.

\begin{table*}
\begin{center}
\caption{Open Cluster parameters and abundances.}
\begin{tabular}{llrrrrrrr}
\hline \hline
 \multicolumn{1}{l}{Id} &
 \multicolumn{1}{c}{Age (Gyr)} &
 \multicolumn{1}{c}{M$_{TO}$ (M$_{\odot}$)$^a$} &
 \multicolumn{1}{c}{R$_{\rm GC}$} (kpc) &
 \multicolumn{1}{c}{[Fe/H]} &
 \multicolumn{1}{l}{12+$\log$~(O/H)} &
 \multicolumn{1}{c}{12+$\log$~(N/H)} &
 \multicolumn{1}{l}{n. stars} &
 \multicolumn{1}{l}{\sc DR} \\
\hline \hline
 Rup134          &   1.00$\pm$0.20     & 2.18 &   4.60$\pm$0.10    &+0.26$\pm$0.06 &   8.95$\pm$0.05   &   8.53$\pm$0.05   &      14    &   {\sc dr5}    \\
  Be81              &   0.86$\pm$0.10  & 2.27    &   5.49$\pm$0.10    &+0.22$\pm$0.07 &   8.94$\pm$0.16   &   8.54$\pm$0.08   &      14    &   {\sc dr4}    \\
  NGC6005      &   1.20$\pm$0.30     & 2.02 &  5.97$\pm$0.34   &+0.19$\pm$0.02 &    8.86$\pm$0.04   &   8.40$\pm$0.04   &     9     &   {\sc dr4}     \\
  Trumpler23    &   0.80$\pm$0.10     & 2.34 &   6.25$\pm$0.15  &+0.21$\pm$0.04 &    8.85$\pm$0.08   &   8.40$\pm$0.08   &     10    &   {\sc dr4}    \\
  NGC6705      &   0.30$\pm$0.05     & 3.30  &  6.33$\pm$0.16  &+0.16$\pm$0.04  &    8.70$\pm$0.03   &   8.51$\pm$0.30     &     16    &   {\sc dr5}    \\
  NGC6067      &   0.10$\pm$0.05     & 5.10  &  6.81$\pm$0.12  &+0.20$\pm$0.08  &    8.88$\pm$0.11   &   8.69$\pm$0.09    &     9     &   {\sc dr5}   \\ 
  Pismis18        &   1.20$\pm$0.04  & 2.03    &  6.85$\pm$0.17  &+0.22$\pm$0.04  &     8.77$\pm$0.06   &   8.43$\pm$0.07   &      4     &   {\sc dr4}    \\
  Be44               &   1.60$\pm$0.30 & 1.85 &     6.91$\pm$0.12  &+0.27$\pm$0.06  &     9.05$\pm$0.20    &   8.39$\pm$0.17    &   7     &   {\sc dr4}    \\
  NGC4815      &   0.57$\pm$0.07     & 2.60 &  6.94$\pm$0.04 &+0.11$\pm$0.01  &     8.80$\pm$0.09   &   8.40$\pm$0.07   &      5     &   {\sc dr2-3}   \\
  NGC6802      &   1.00$\pm$0.10     & 2.12 & 6.96$\pm$0.07 &+0.10$\pm$0.02  &     8.71$\pm$0.14   &   8.35$\pm$0.06   &     9     &   {\sc dr4}  \\   
  NGC6259      &   0.21$\pm$0.03     & 3.88 &  7.03$\pm$0.01 &+0.21$\pm$0.04  &     8.86$\pm$0.05   &   8.62$\pm$0.05   &     9     &   {\sc dr5}    \\
  NGC6633      &   0.52$\pm$0.10     & 2.63 & 7.71$\pm$0.01 &-0.01$\pm$0.11   &     9.01$\pm$0.28   &   8.20$\pm$0.30  &     11    &   {\sc dr5}    \\
  M67               &   4.30$\pm$0.50 & 1.30     &   9.05$\pm$0.20  &-0.01$\pm$0.04   &    8.61$\pm$0.09   &   8.00$\pm$0.05   &   12    &  {\sc dr5}    \\
  NGC2243      &   4.00$\pm$1.20   & 1.20  & 10.40$\pm$0.20   & -0.38$\pm$0.04 &   8.57$\pm$0.07   &   7.57$\pm$0.17   &     8     &   {\sc dr5} \\   
  Melotte71      &   0.83$\pm$0.18   & 2.18 &  10.50$\pm$0.10    & -0.09$\pm$0.03 &  8.65$\pm$0.04   &   8.06$\pm$0.08  &     4     &   {\sc dr5}   \\ 
  Be36              &   7.00$\pm$0.50   &  1.06 & 11.30$\pm$0.20    &  -0.16$\pm$0.10 & 8.80$\pm$0.02   &   8.00$\pm$0.05   &     5     &   {\sc dr5}    \\
  Be31              &   2.50$\pm$0.30   &  1.44 & 15.20$\pm$0.40    &   -0.27$\pm$0.06& 8.65$\pm$0.03   &   7.77$\pm$0.09   &     7     &   {\sc dr5}    \\
 \hline\hline
\end{tabular}
\label{tab_clu}
\end{center}
(a) M$_{TO}$ derived from Parsec isochrones using ages and [Fe/H] in the table
\end{table*}

\section{The effect of mixing}
\label{sec:mixing}

\begin{figure}
   \centering
  \includegraphics[width=.5\textwidth]{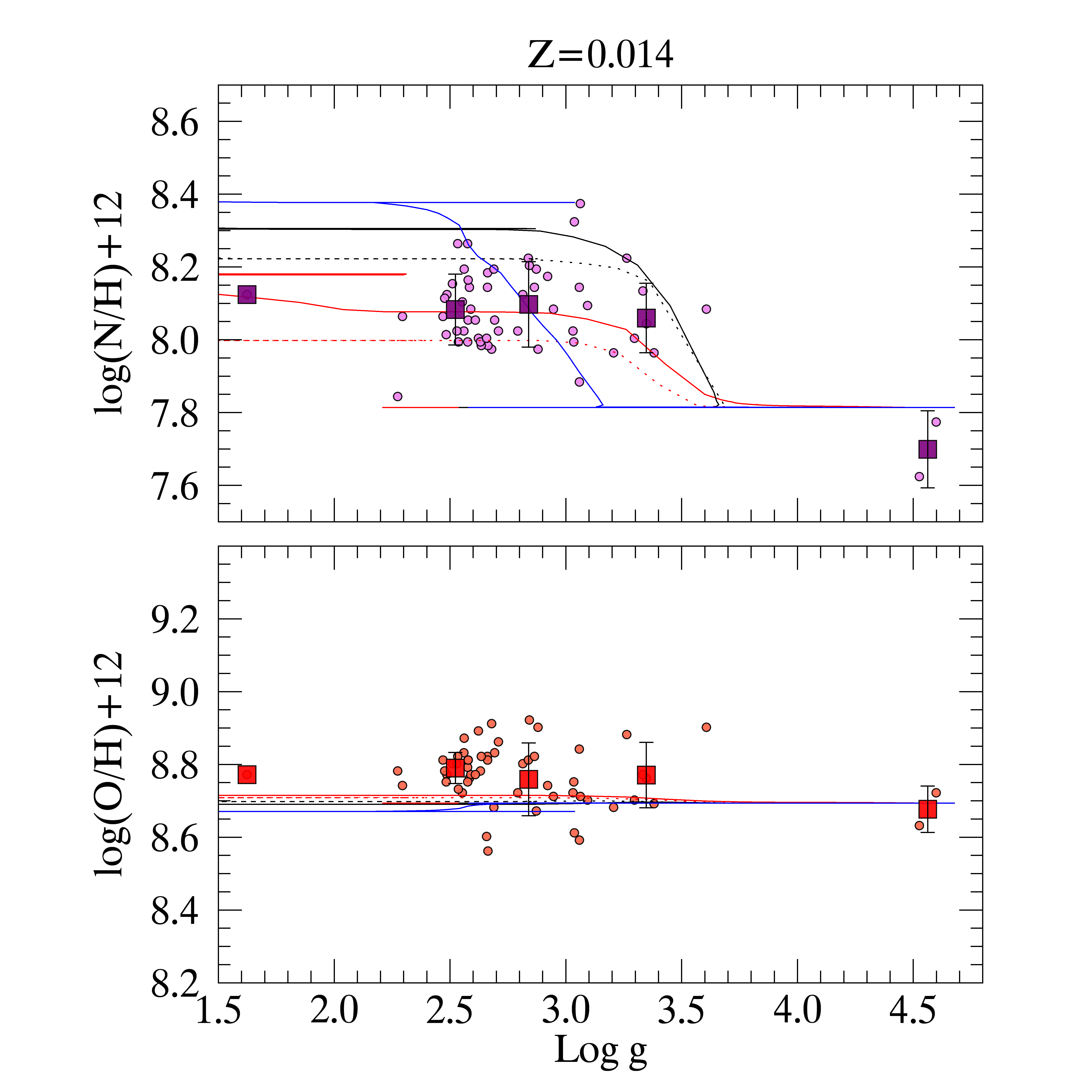}
  \caption{12+$\log$(O/H) and 12+$\log$(N/H) versus surface gravity in the field and cluster stars (circle are single measurements, while with squares are the results binned 
  in $\log$~g bins of 0.5. The curves are the model of \citet{lagarde12} for 1~M$_{\odot}$ (red dashed-ST, red continuous-TCR), for 2~M$_{\odot}$ (black   dashed-ST, black continuous-TCR) and for 3~M$_{\odot}$ (blue continuous-TCR).  }
        \label{mixing}
\end{figure}

Stellar evolution can affect the abundances of C and N, and thus they do not trace the initial composition of the stars. On the other hand,    
the abundances of O should reflect the chemical composition of the stars at their birth \citep{tau15}. 
Using a set of stellar evolution models with both thermohaline and rotation induced mixing by \citet{lagarde12}, we estimate the effect of these processes 
on the measured abundances of N and O. 
In Figure~\ref{mixing}, we show the abundances of N and O as a function of surface gravity for both stars in clusters and field in the metallicity range -0.05$\leq$[Fe/H]$\leq$+0.05.
We compare the data with a set of models by \citet{lagarde12} computed for solar metallicity with both standard prescriptions (ST) and  with thermohaline convection and rotation-induced mixing (TCR). We plot the models for three stellar masses: 1, 2 and 3 M$_{\odot}$. The mass range 1~M$_{\odot}$$\leq$M$\leq$3~M$_{\odot}$ covers, indeed,  most of the lifetime of the thin and thick disc stars, corresponding to approximately a time interval $\sim$0.6-10~Gyr. 
From the upper panel of Figure~\ref{mixing}, nitrogen is enhanced during the latest phases of the stellar evolution: N/H in  the few dwarf stars with surface gravities $\log$~g$\sim$4.5 is lower than N/H in giant stars. For the field stars, the enhancement in N/H is reproduced considering the models with 1~M$_{\odot}$$\leq$M$\leq$2~M$_{\odot}$.

In the bottom panel of Figure~\ref{mixing}, we show oxygen abundances versus surface gravity. In this case, the effect of mixing is negligible, both from models and observations. Thus the measured oxygen abundance can be considered representative of the initial composition of the stars.   We note, however, a relevant spread in the O abundances possibly related to two different aspects: the presence of both thin and thick disc stars, the latter more enhanced in O/Fe with respect to the thin disc stars; the missing correction of telluric absorption which is not included in the Gaia-ESO standard reduction.

\begin{figure}
   \centering
  \includegraphics[width=.5\textwidth]{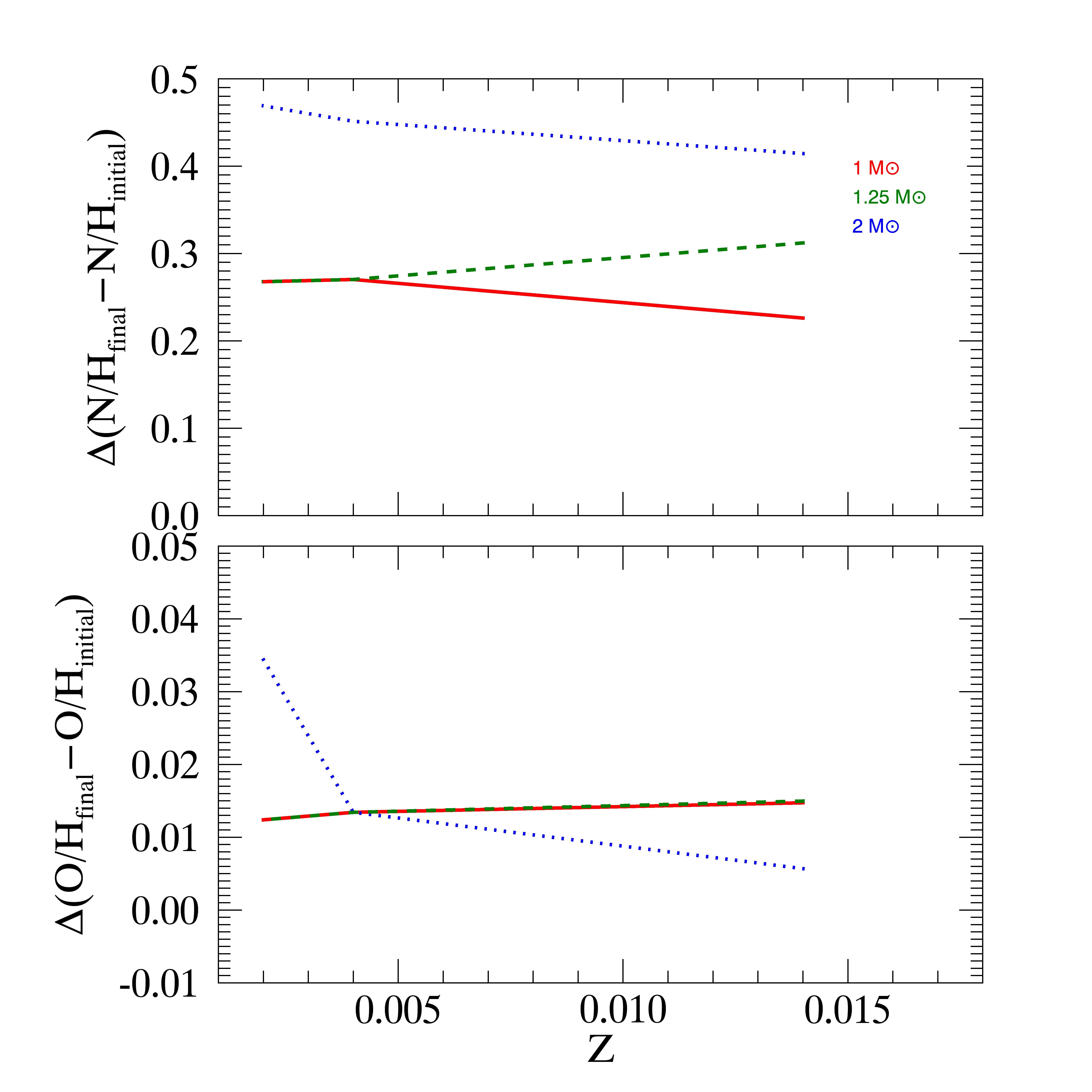}
  \caption{$\Delta_{\rm N/H}$=N/H$_{final}$-N/H$_{initial}$ and $\Delta_{\rm O/H}$=O/H$_{final}$-O/H$_{initial}$ versus Z metallicity in the ST models of \citet{lagarde12} 
  for stars of 1, 1.25 and 2 M$\odot$.  }
        \label{fig_correction}
\end{figure}

We have also investigated the effect of metallicity in the enhancement of nitrogen (and of oxygen) during stellar evolution to see if we are allowed to apply, in first approximation, an average correction to our stars, which span a metallicity range of about $\sim$1~dex. 
In Figure~\ref{fig_correction}, we show the variation of surface abundances of nitrogen and oxygen at different metallicities for three representative stellar masses 
(1, 1.25 and 2 M$\odot$) using the ST models of  \citet{lagarde12}. 
In our metallicity range (0.005$\leq$Z$\leq$0.015) there are no strong variations of $\Delta_{\rm N/H}$=N/H$_{initial}$-N/H$_{measured}$ and an average value of -0.25~dex is a good approximation for low mass stars
(mostly representative of the field thin and thick disc populations). 

\begin{figure}
   \centering
  \includegraphics[width=.5\textwidth]{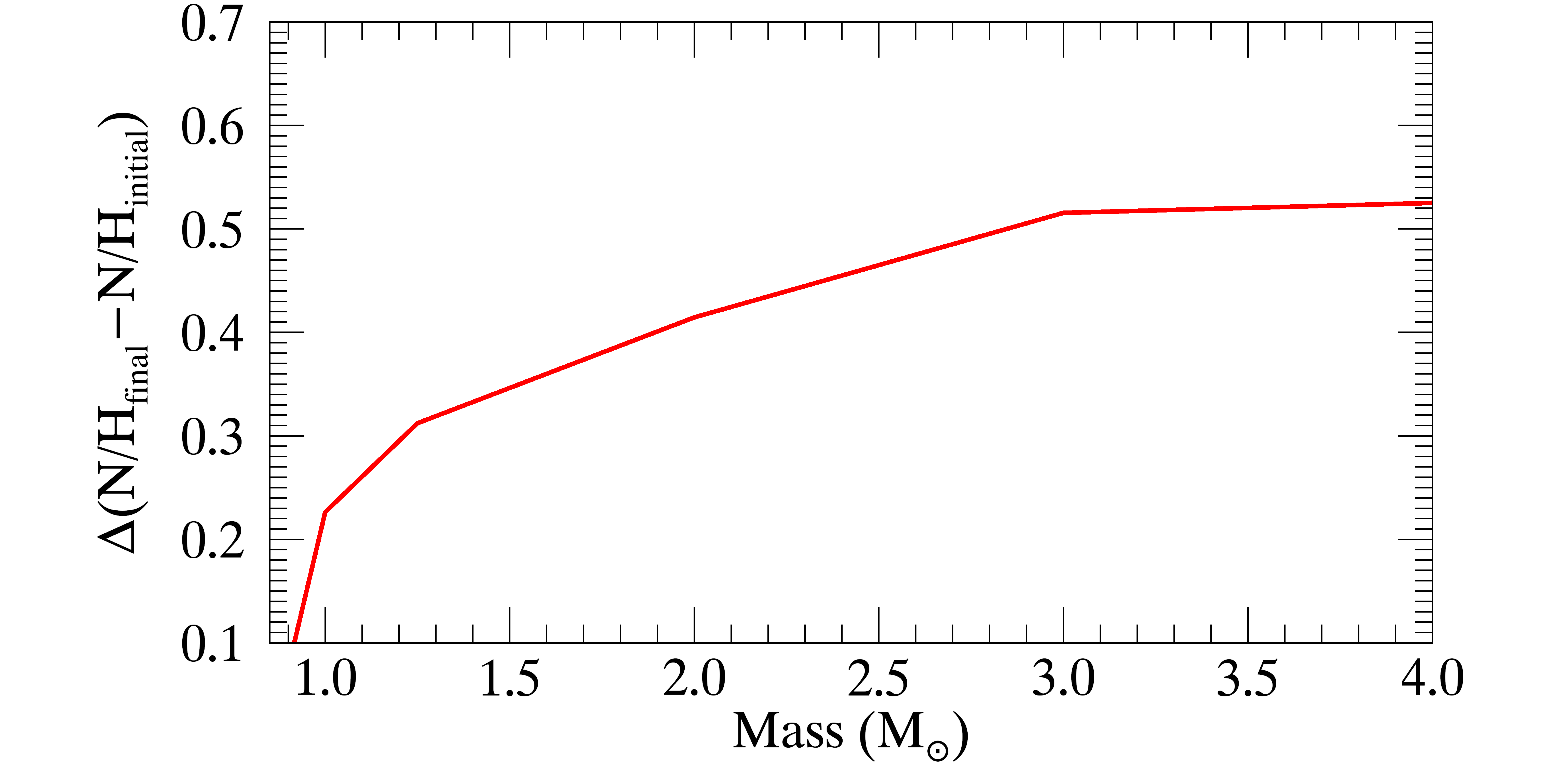}
  \caption{$\Delta_{\rm N/H}$=N/H$_{final}$-N/H$_{initial}$ versus mass in the ST models of \citet{lagarde12} at solar metallicity. }
\label{fig_correction_m}
\end{figure}

On the other hand, for the open clusters, the average correction valid for field stars is, in general,  an under-estimation since they are younger and thus their evolved stars are more massive. 
Using the mass at turn-off of clusters uniformly derived Parsec isochrones \citep{parsec}
with the ages and metallicities of Table~2, we estimate the variation of nitrogen surface abundance during stellar evolution using the models of \citet{lagarde12} for more massive stars. In Figure~\ref{fig_correction_m} we plot $\Delta_{\rm N/H}$=N/H$_{initial}$-N/H$_{measured}$ as a function of stellar mass for solar metallicity. 
The variation $\Delta_{\rm N/H}$ is larger for more massive stars and it can be approximated with a polynomial fit. 
A fit to the data of Figure~\ref{fig_correction_m} gives us the possibility to correct the N/H abundances for giant stars in clusters for which we know the turn-off mass. 
The fit is given in the following equation, where $m$ is the turn-off mass: 
\begin{equation}
\Delta_{\rm N/H}=-(-2.5+ 5.4\times m-3.5\times m^2+0.99\times m^3-0.09\times m^4)
\end{equation}

In the following plots and discussions, we have applied a correction to 12+$\log$(N/H) abundances and to N/O abundance ratios, both in clusters and field stars. For the field population, we have considered a constant correction of -0.25~dex, while for the cluster stars we have used the fit in Eq.~1 using the turn-off masses of Table~2. The errors due to this correction are difficult to estimate and they are not propagated in the final abundances.  
In the following analysis, for both populations we have considered only stars with $\log$g$\leq$3.5~dex, which have more reliable determination of O and N abundances.

\section{The abundance results}

\begin{figure}
   \centering
  \includegraphics[width=.5\textwidth]{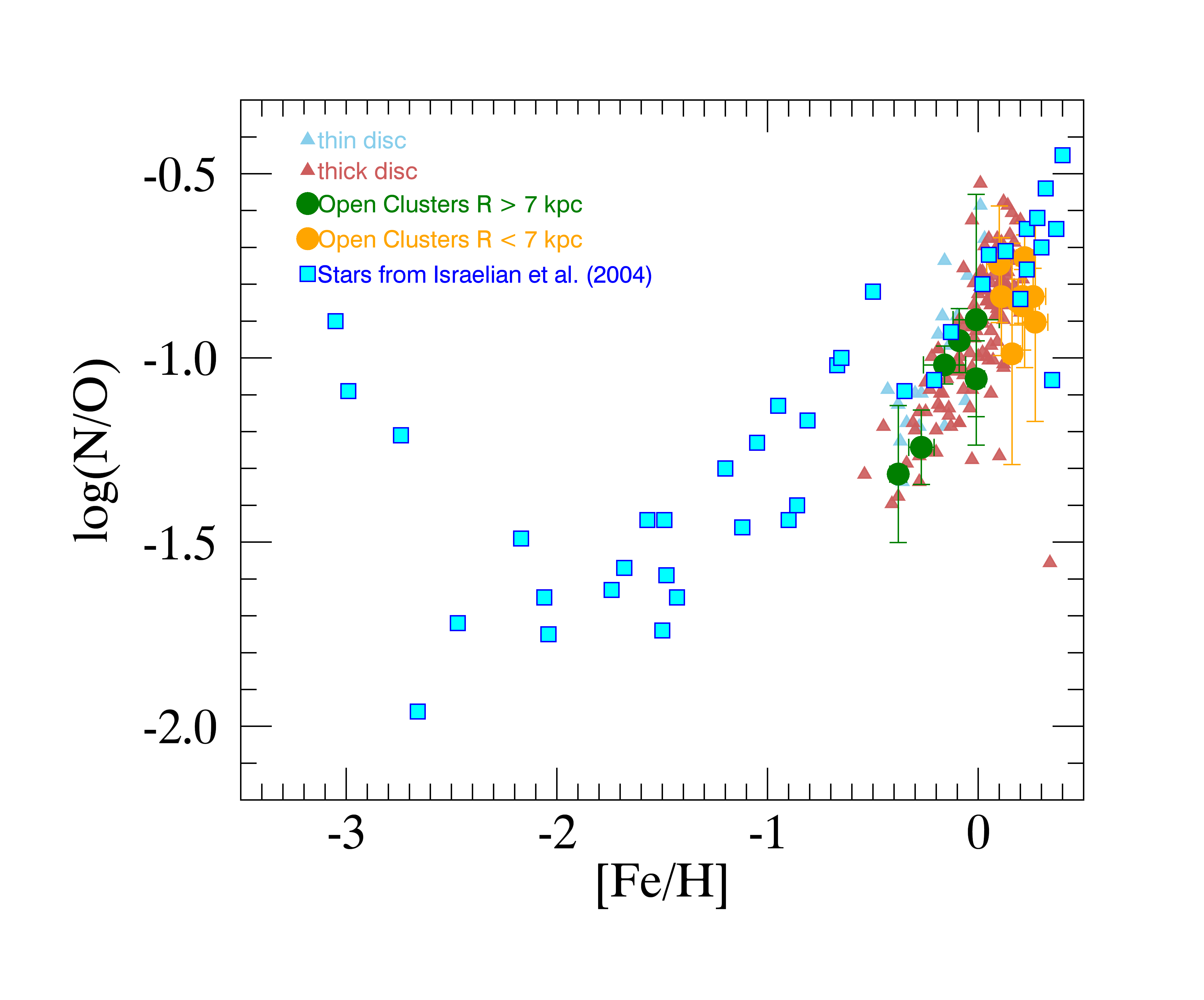}
  \caption{$\log$(N/O) versus [Fe/H] in our samples compared with the literature data of \citet{israelian04}. Open clusters are shown with large filled circle (in green clusters with R$_{\rm GC} >$7~kpc and in orange clusters with R$_{\rm GC} \leq$7~kpc), thin and thick disc stars are represented with smaller light blue and pink triangles, respectively. The literature sample of \citet{israelian04} is represented with cyan squares. }
\label{literature}
\end{figure}

In Figure~\ref{literature} we compare our results, after correcting nitrogen abundances for the effects of stellar evolution,  with those collected by \citet{israelian04} for both  metal-poor and metal-rich dwarf stars. 
On the one hand, in the metallicity interval -0.5$<$[Fe/H]$<$0.5 spanned by our new data, there is a very good agreement between the GES and the literature results. 
On the other hand, the sample of \citet{israelian04} reaches very low metallicities, not available in our sample dominated by the disc populations. 

\begin{figure}
   \centering
  \includegraphics[width=.5\textwidth]{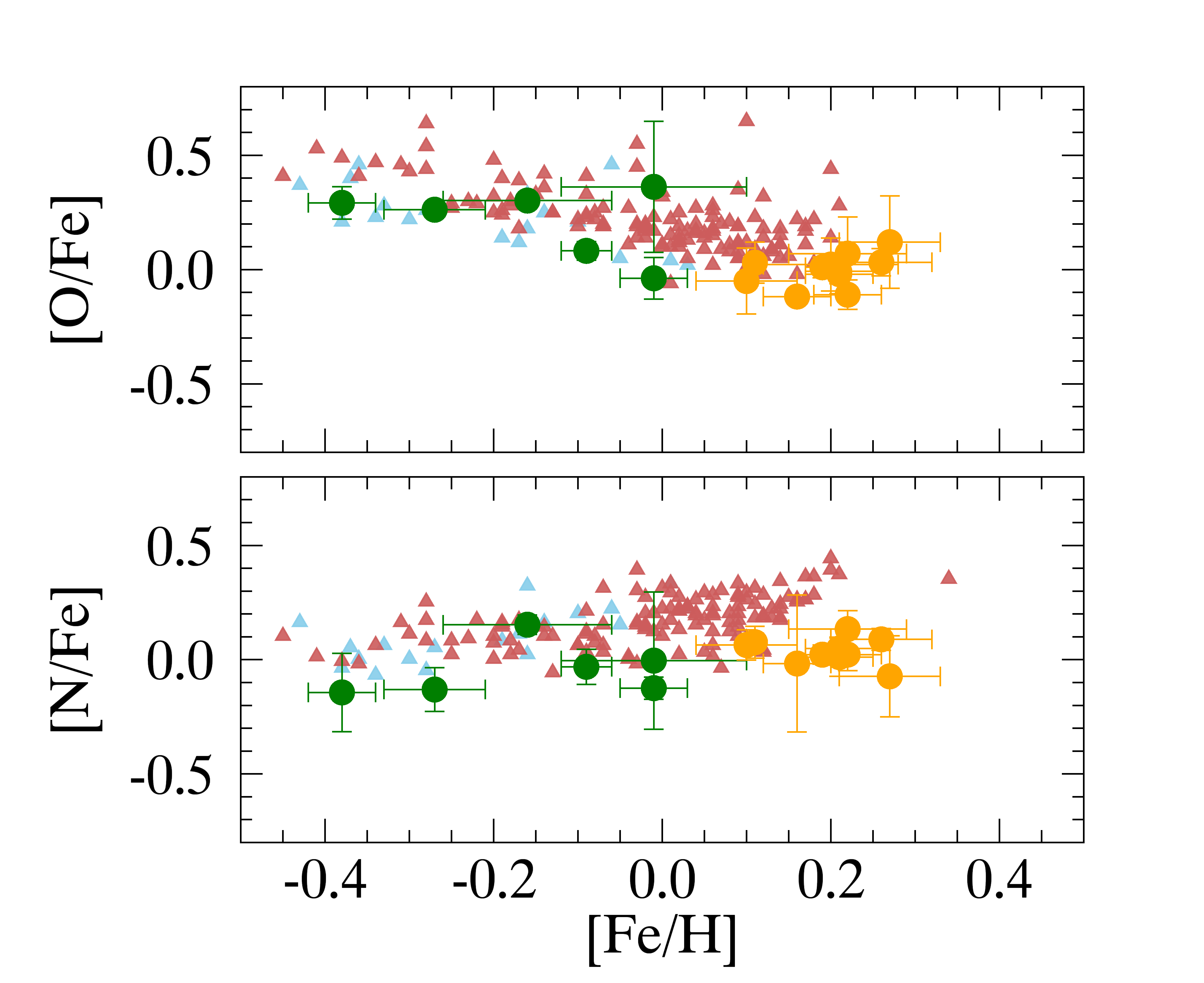}
  \caption{[N/Fe] and [O/Fe] as function of [Fe/H]  in the thin and thick disc populations and in open clusters. Symbols and colour codes 
  as in Figure~\ref{literature}. }
\label{nfe_ofe}
\end{figure}

In Figure~\ref{nfe_ofe}, we show [N/Fe] and [O/Fe] as function of [Fe/H]  in the thin-, thick-disc populations and in open clusters. 
We performed the two-dimensional Kolmogorov-Smirnov (KS) test \citep[][]{fasano87} to quantify the probability that the abundance ratios in the three populations derive from similar distributions. 

The bi-dimensional KS  statistical test is the generalization of the classical one-dimensional KS test and it is used to analyse two- or three-dimensional samples.
Based on the two-dimensional KS test, the significance of the equivalence between the distributions of [O/Fe] and of [N/Fe] in thin disc stars with thick disc stars is less than 1\% in the overlapping metallicity region. 
On the other hand, the abundance ratios [N/Fe] and [O/Fe] in open clusters are much similar to those in thin disc stars with probabilities ranging from $\sim$20\% to $\sim$50\%, respectively.

\section{The chemical evolution models}
\label{sec:models}

We compare our observed data sample with the predictions of classical one-zone chemical evolution models, making a very similar approach 
as in \citet{vincenzo16}, with standard assumption that are used to reproduce the main characteristics of our Galaxy. The parameters of our reference model are in agreement with many previous studies \citep[e.g.][]{minchev2013,spitoni2015,spitoni2017}. 

In summary, the star formation rate (SFR) is assumed to follow a linear Schmidt-Kennicutt law, namely 
$\text{SFR}(t)=\text{SFE}\times M_{\text{gas}}(t)$, where SFE represents the star formation efficiency, a free parameter of our models, and $M_{\text{gas}}(t)$ is the 
total amount of gas in the galaxy at the time $t$. 
We assume the Galactic disc to assemble by accreting primordial gas 
with a rate exponentially decaying as a function of time, namely $I(t) \propto \exp(-t/\tau)$, 
normalised such that $ \int_{0}^{t_{\text{G}}}{dt\,I(t)} = M_{\text{inf}}$, 
with $\tau$ representing the so-called ''infall time scale'', a free parameter of our models, and $M_{\text{inf}}$ 
the integrated amount of gas accreted into the Galactic potential well during the Galaxy lifetime ($t_{\text{G}}=14\,\text{Gyr}$); in this work, 
we assume $\log( M_{\text{inf}}) = 11.5\,\text{dex}$. 

In this work, the effect of galactic winds is included, by assuming that the outflow rate is directly proportional to the SFR, namely  $O(t) = w \times \text{SFR}(t)$, where 
$w$ represents the so-called mass loading factor, another free parameter of our models. 
The wind is differential and it is assumed to carry only the main nucleosynthetic products of Supernovae (SNe), hence $\alpha$- and iron-peak elements. 
To compute the time of onset of the galactic wind, we follow the same formalism as in \citet{bradamante1998}. Finally, we assume the initial mass function (IMF) of \citet{salpeter1955}. 

For LIMS, 
we assume the same stellar yields as in \citet{vincenzo16}, while for massive stars  the stellar yields of \citet{kobayashi2011}. 
In order to reproduce the low-metallicity N/O plateau, we assume that massive stars only produce primary N, 
with an empirical stellar yield which is computed as in \citet{vincenzo16}. For Type Ia SNe, we assume the 
following delay time distribution function (DTD): $\text{DTD}_{\text{Ia}}(t)\propto 1/t$ \citep{totani2008}, which gives very similar results 
as the DTD of \citet{schoenrich2009}, and nucleosynthetic yields from 
\citet{iwamoto1999}. The assumed DTD is normalised to have a total number of SNe in the range $\sim1$-$2$ SNe per $10^{3}\,\text{M}_{\sun}$ of stellar mass 
formed for all our models \citep{bell2003,maoz2014}. Finally, we assume the metallicity-dependent stellar lifetimes of \citet{kobayashi2004}. 

We have made several numerical experiments, by assuming different 
prescriptions for Type Ia SNe, IMF and stellar yields, and we have came to the conclusion that the aforementioned assumptions provide the best match to the 
observed data sample for [O/Fe] versus [Fe/H] and 
$\log$(N/O) versus [Fe/H]. 

We construct a grid of chemical evolution models by continuously varying the free parameters in the following ranges: 
$0.2\le \text{SFE}\le 4\,\text{Gyr}^{-1}$, $1\le \tau \le 12\,\text{Gyr}$ and $0.4\le w \le 1.0$. In the next Section, when we refer to our 
reference ''grid of models'', we mean the aforementioned variation of the free parameters. Our reference chemical evolution model assumes 
$\text{SFE} = 1\,\text{Gyr}^{-1}$, $\tau = 7\,\text{Gyr}$ and $w = 0.8$, which can reproduce the observed average [O/Fe]--[Fe/H] and 
N/O--[Fe/H] abundance patterns. 

The main differences between the chemical evolution models of \citet{vincenzo16} and those of this work can be summarised as follows. \textit{(i)} Here we assume the IMF of \citet{salpeter1955}, containing a lower number of LIMS than the \citet{kroupa1993} IMF assumed in \citet{vincenzo16}. \textit{(ii)} We assume the double degenerate scenario for Type Ia SNe, while \citet{vincenzo16} assume the single-degenerate scenario of \citet{matteucci2001}.

The main limitation of our approach is that we adopt a one-zone model to characterise the whole disc of our Galaxy, varying the main free parameters to reproduce the observed chemical abundance patterns. There are chemical abundance gradients in our Galaxy \citep[see, e.g.][]{esteban2018}, which can be seen also in our dataset and can be better understood -- from a physical point of view -- only by making use of chemodynamical simulations embedded in a cosmological framework \citep[e.g.][]{kobayashi-nakasato2011, vincenzo18}. Finally, we assume in our model a single onset for the galactic wind, which is maintained continuously afterwards; this may not be entirely appropriate for actively star forming galaxies which may experience large variations in their SFRs. 

\section{Results} 

\begin{figure}
   \centering
  \includegraphics[width=0.5\textwidth]{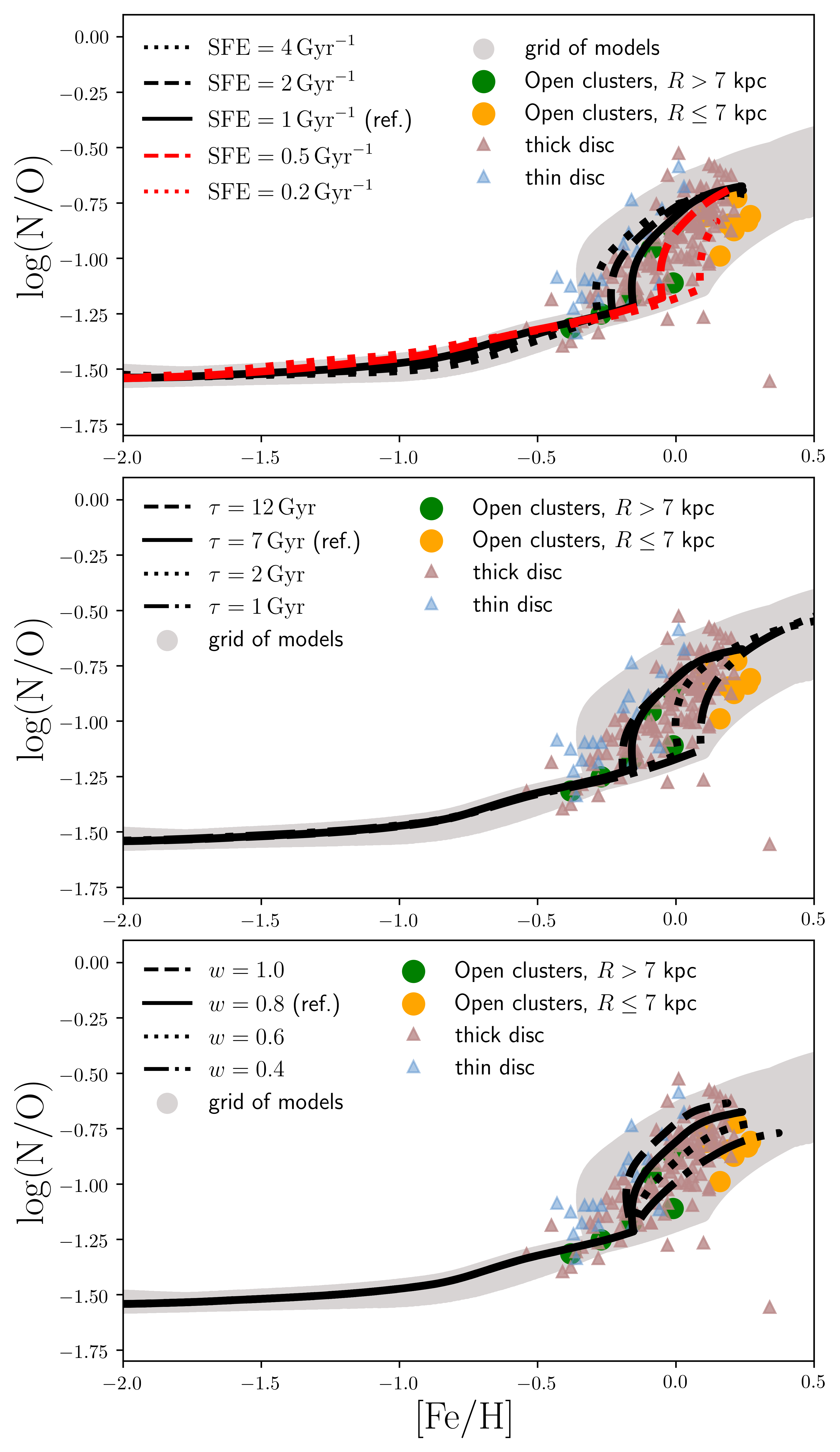}
  \caption{$\log$(N/O) versus [Fe/H] with open clusters shown as large filled circle (in green clusters with R$_{\rm GC} >$7~kpc and 
  in orange clusters with R$_{\rm GC} \leq$7~kpc), individual measurements of thin and thick disc stars are represented with cyan and pink triangles, respectively. 
  In the top panel, we present the grid of models in which we vary the SFE; in the central panel, the infall time scale is changed, while in the bottom panel,  the outflow loading factor is varied. In each panel, the reference model is identified with
  a solid line.  }
        \label{models}
\end{figure}

In Figure~\ref{models}, we compare $\log$(N/O) versus [Fe/H] of field stars (divided in thin and thick disc) and the open clusters with the set of chemical evolution models from our grid. 
In all panels we also report the reference model. 
First we note the clusters and stars of the Milky Way are located in the upper-right part of the plot, where the secondary production of N dominates. 
Conversely to studies of unresolved galaxies, with the stellar population of the Milky Way we can appreciate the spanned ranges of metallicities and of $\log$(N/O) belonging to the same galaxy. 
Thanks to the precise  measurements of distances of open clusters we can also relate the variation of N/O to different parts of the Galaxy. 


In the top panel, we vary the SFE, from 0.2~Gyr$^{-1}$  to 4~Gyr$^{-1}$.  
As discussed in \citet{vincenzo16}, the effect of varying SFE is to increase the metal content and thus to move the point where the secondary component of N starts to contribute.  As aforementioned, since we assume an IMF which is not bottom-heavy \citep{salpeter1955}, the N production from LIMS is less prominent than in \citet{vincenzo16}. The onset of the galactic wind in the models is crucially determined by the interplay between the injection rate of thermal energy by SN events and stellar winds and the assembly of the Galaxy potential well as a function of time (mostly regulated by the assumed infall mass and infall timescale). By lowering the SFE in the model (keeping fixed the other parameters), there is less thermal energetic feedback from SNe and stellar winds at any time of the galaxy evolution, delaying the onset of the galactic wind towards higher [Fe/H] abundances. We remark on the fact that the galactic wind onset in Figure~\ref{models} appears as a sudden increase in $\log$(N/O), from a given [Fe/H] on. 
To reproduce the range of $\log$(N/O) spanned by the observations, a variety of SFE is necessary. 
In particular, to reproduce the observed abundances in the thin disc stars, 
we need higher average SFEs than for thick disc stars. 
Here and in the following discussion we consider that young and intermediate-age open clusters are not affected by strong radial migration
and thus they are representative of the abundances of the place where they are observed \citep[cf., e.g.,][]{quillen18}.


In the central panel, we vary the infall timescale, from 1~Gyr  to 10~Gyr keeping the SFE fixed at its reference values. 
The main effect of the infall timescale is that models with longer infall time-scales have a galactic wind that
develops at earlier times, and thus the location of the break point due to the onset of the wind (see Figure~3 of V16) is moved and the slopes of the models after the break point are changed. 
The Milky Way resolved data are better reproduced by models with longer infall timescales in agreement with other evidence of time scales of the order of $\sim$8~Gyr of the Galactic thin disc at Solar Galactocentric distance.  
The inner disc open clusters can be  better reproduced with models with shorter infall time scales, while the outer 
disc open clusters need a longer infall time scale  in agreement with the inside/out scenario.  The assumption of infall timescales $\tau_{\text{inf}}>7$~Gyr has little effect on the chemical evolution tracks in Figure~\ref{models}, for a fixed SFE, confirming that the main parameter in our model to discriminate between thin and thick disc stars is given by the SFE.  

In the bottom panel, we vary the effects of differential outflow, the so-called outflow loading factor. In our models, we assume a differential outflow, which carries out 
only the main nucleosynthetic products of SNe, thus affecting oxygen, but not nitrogen. 
In the plot, we vary it from $\omega$=0.4  to $\omega$=1.0, while the reference model has $\omega$=0.8. 
In the case of the Milky Way, a  $\omega$=1.0 better suited to reproduce the high metallicity data of field stars,  while for the inner disc clusters a lower loading factor is needed (see Figure~\ref{models}, bottom panel). This may be justified by the fact that the ISM in innermost regions of our Galaxy is more tightly bound than in the outermost ones, being eventually ejected with higher average efficiencies.

Although our model might be simplistic, it can capture some of the main features of the observed chemical abundance data. We remark on the fact that, by moving along each chemical evolution track in Figure~\ref{models}, the SFR and hence the predicted number of stars with a given  $\log$(N/O) and [Fe/H] can vary; hence there is a third hidden dimension in the tracks of Figure~\ref{models}. 
For example, the model with the lowest SFE in the top panel contains very few stars after the onset of the galactic wind; conversely, the model with the shortest infall timescale in the middle panel contains a large number of stars also after the development of the galactic wind. Therefore, even though SFE and infall timescale seem to suffer from some degeneracies in the chemical evolution tracks of Figure~\ref{models}, their relative contributions might be better discriminated by looking also at the number of stars with a given $\log$(N/O) and [Fe/H] in the model and in the data. Nevertheless, to do this kind of study, we need a much larger and more complete statistical dataset of chemical abundances in our Galaxy than the one presented by this work. Large datasets are currently available, for example from APOGEE, GALAH, and LAMOST \citep{apo, galah, lamost}, however, they suffer from lower resolution than in this work. 

 Putting together all these considerations, we can conclude that the reference model alone is not able to reproduce the whole abundance ratios along the radial range of the disc. Due to radial variations of the Galactic properties and to the inside-out formation of the disc, this is indeed expected.  
The more efficient way to reproduce the abundance ratios in the outermost clusters is by assuming relatively long infall timescales ($\tau_{\text{inf}}\sim 7$-$8$~Gyr) and SFEs of the order of unity per Gyr, which is perfectly in line with the inside-out scenario for the formation of the MW disc. We find that thin disc stars tend to have higher $\log(\text{N/O})$ at fixed $\log(\text{O/H})+12$ than thick disc stars; this can be reproduced only by assuming for thin disc stars higher average SFEs and longer infall timescales, with the effect of the SFE being more important than the effect of the infall timescale. Finally, lower outflow loading factors seem also to better reproduce the abundance patterns in the inner open clusters. 

\section{The Milky Way in the extragalactic framework}
\begin{figure*}
   \centering
  \includegraphics[width=0.95\textwidth]{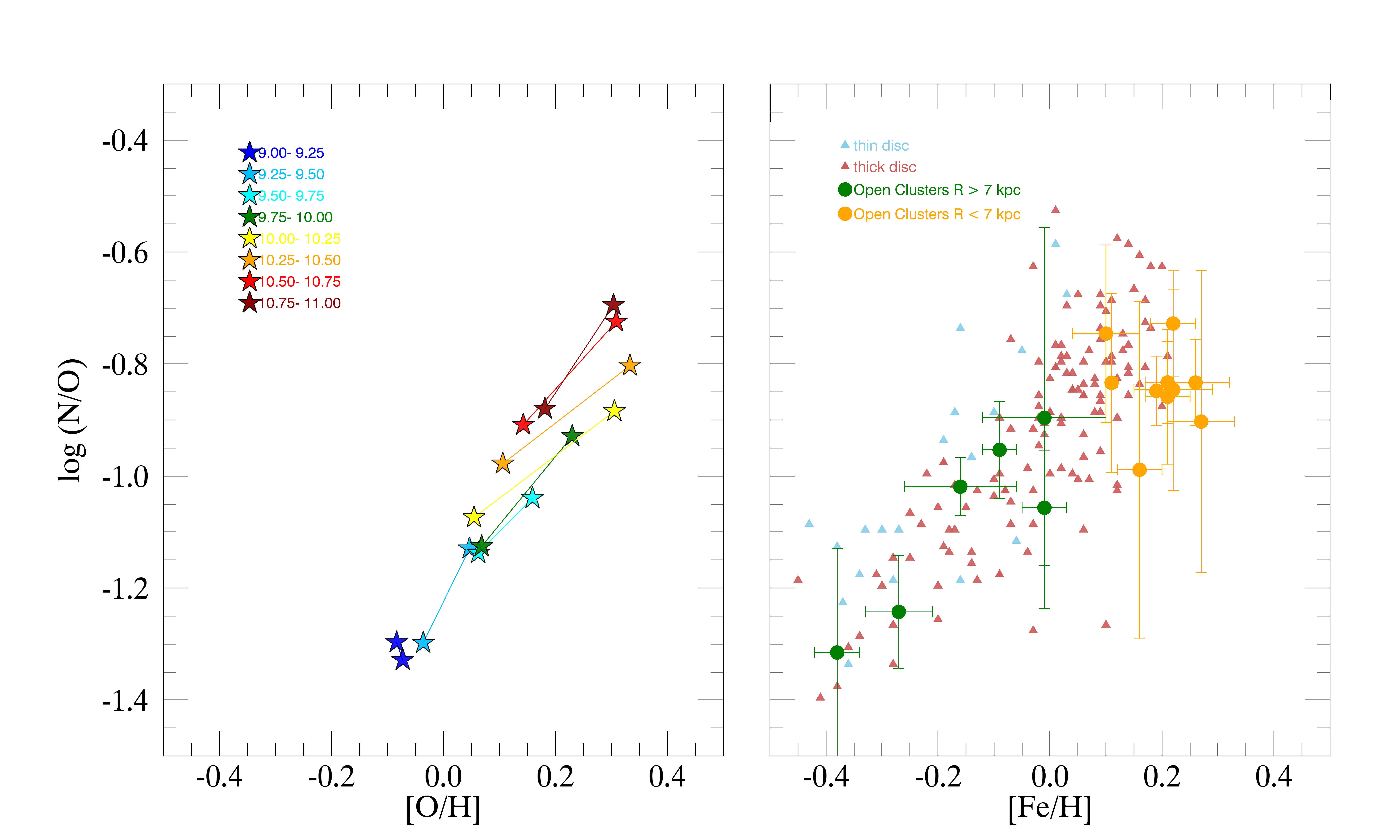}
  \caption{Left panel: $\log$(N/O) vs [O/H] as a function of both stellar mass and radius in the sample of resolved galaxies of \citet{belfiore17} (extracted from their Figure 8): for each mass bin (see the legend for the range in logarithm of the stellar mass of each bin) the the uppermost star represents 
   the innermost radial bin (0.0-0.25 R$_{\rm e}$ while the lower star represents the outermost radial bin (1.75-2.0 R$_{\rm e}$).
  Right panel: $\log$(N/O) versus [Fe/H] in our Milky Way samples of field stars (blue and pink triangles, thin and thick discs, respectively) and of open clusters (in green clusters with R$_{\rm GC} >$7~kpc and 
  in orange clusters with R$_{\rm GC} \leq$7~kpc).    }
        \label{comparison}
\end{figure*}

In Figure~\ref{comparison}, we compare our results (right panel) with the MaNGA sample \citep{belfiore17} 
that includes 550  resolved galaxies with stellar masses between 10$^{9}$ and 10$^{11}$ M$_{\odot}$ (left panel).
The Milky Way stellar data span an interval in $\log$(N/O) as large as the whole range spanned by external galaxies belonging to different bins of stellar masses.
The stellar mass of our Galaxy is estimated to be 6.1$\pm$1.1$\times$10$^{10}$M$_{\odot}$ \citep{licquia15}.  
The bulk of the thick disc stars and the innermost clusters are indeed in agreement with $\log$(N/O) of external galaxies in the stellar mass bin 10.50-10.75. 
On the other hand, the  outer regions of the Milky Way show $\log$(N/O) values consistent with those of  low stellar mass galaxies, in which the SFE is typically lower. 
In addition, in the very inner clusters and in high metallicity thick disc stars, we observe the highest values of $\log$(N/O), reached {\bf in inner parts} of the more massive star forming galaxies. 
 Such high values are also reached  in resolved H~{\sc ii} regions of nearby galaxies (see, e.g., Fig.~6 of \citet{bresolin04} where measurements in M51, M101 and NGC2403 are shown). 
The radial variation of $\log$(N/O) in the Galaxy  can be interpreted as a consequence of the different SFE, infall time-scales and galactic wind between the central and the outer regions of discs, 
and not only the H{\sc II} regions but also Galactic stellar populations keep trace of the 
different mechanisms of formation of different parts of the MW. 



\section{Summary and conclusions} 
We present the Gaia-ESO {\sc idr2-3, idr4, idr5} abundances of N and O  in open clusters and in field stars belonging to thin and thick discs. 
We estimate the effect of stellar evolution on N abundance, and we correct the abundance of giant stars making used of the stellar evolution models 
of \citet{lagarde12}. 
The study of individual stars and clusters has shown 
that the stellar populations of the MW present a  wide range of $\log$(N/O), mainly located in the region dominated by the secondary production of N. This distribution is comparable to their range observed in individual  H{\sc II} regions  in external massive galaxies, or even larger \citep[e.g.][]{bresolin04}. 
We compare  $\log$(N/O) versus [Fe/H] in our samples with a grid of chemical evolution models, based on the models  of \citet{vincenzo16}. 
The Galactic  reference model alone is not able to reproduce all the resolved Galactic trends: we need a higher SFE and/or a longer infall time-scale in the outskirts to reproduce the abundance ratios in the outer regions, coupled with a shorter infall time-scale in the inner regions together with a lower outflow loading factor to explain  the abundance ratios in the inner disc. 
This is in line with the request of an inside-out formation in the MW disc. Interestingly, thin disc stars have higher average $\log(\text{N/O})$ ratios at fixed $\log(\text{O/H})+12$ than thick disc stars; this feature can better reproduced by models assuming higher average SFEs for thin disc stars. 
We also compared our resolved Milky Way results with the local Universe sample of MaNGA, finding a remarkably good agreement, but with the MW data spanning a larger range of $\log$(N/O) than integrated observations of galaxies of similar stellar mass.


    \begin{acknowledgements}
We thank the referee for her/his constructive report that improved the quality and presentation of the paper. We thank Viviana Casasola for her help the KS bi-dimensional test. 
Based on data products from observations made with ESO Telescopes at the La Silla Paranal Observatory under programme ID 188.B-3002. These data products have been processed by the Cambridge Astronomy Survey Unit (CASU) at the Institute of Astronomy, University of Cambridge, and by the FLAMES/UVES reduction team at INAF/Osservatorio Astrofisico di Arcetri. These data have been obtained from the Gaia-ESO Survey Data Archive, prepared and hosted by the Wide Field Astronomy Unit, Institute for Astronomy, University of Edinburgh, which is funded by the UK Science and Technology Facilities Council (STFC).
This work was partly supported by the European Union FP7 programme through ERC grant number 320360 and by the Leverhulme Trust through grant RPG-2012-541. We acknowledge the support from INAF and Ministero dell' Istruzione, dell' Universit\`a' e della Ricerca (MIUR) in the form of the grant "Premiale VLT 2012". The results presented here benefit from discussions held during the Gaia-ESO workshops and conferences supported by the ESF (European Science Foundation) through the GREAT Research Network Programme. FV acknowledges funding from the UK 
STFC through grant ST/M000958/1. 
T.B. was supported by the project grant 'The New Milky Way' from the Knut and Alice Wallenberg Foundation.
A.B. acknowledges support from the Millennium Science Initiative (Chilean Ministry of Economy). 
GT, AD, {\v S}M, RM, ES, and VB acknowledge support from
the European Social Fund via the Lithuanian Science Council grant No.09.3.3-LMT-K-712-01-0103.
\end{acknowledgements}

\end{document}